\documentclass[slac_one]{revtex4}

\usepackage{fancyhdr}

\usepackage{subfigure}
\usepackage[pdftex]{graphicx}
\usepackage{epstopdf}

\newcommand{\ppbar} {\mbox{$p\bar{p}$}}
\newcommand{\qqbar} {\mbox{$q\bar{q}$}}
\newcommand{\ttbar} {\mbox{$t\bar{t}$}}
\newcommand{\met} {\not \!\! E_T}

\newcommand{\gev}{\rm{GeV}}
\begin{document}

\title{Top Quark Pair Production Cross-Section Measurements at D\O }

\author{J. P. Konrath}
\affiliation{On behalf of the D\O\ Collaboration}
\affiliation{Physikalisches Institut der Albert-Ludwigs-Universit\"at Freiburg, Hermann-Herder-Strasse 3a, 79104 Freiburg, Germany}

\begin{abstract}
We present recent measurements of the top quark pair production cross-section in the dilepton, lepton+jets and lepton+tau final states with the D\O\ detector in $\ppbar$ collisions at $\sqrt{s}=1.96$ TeV. %These measurements have been carried out with an integrated luminosity of 1 fb$^{\rm{-1}}$, an exception being the lepton+$\tau$ cross-section measurement.
\end{abstract}

\maketitle

\thispagestyle{fancy}

\section{INTRODUCTION}
The top quark was discovered in 1995 by the D\O\ and CDF collaborations and completes the quark sector of the Standard Model (SM). Its production is expected to occur via strong interactions, namely $\qqbar$ annihilation and gluon fusion. It is expected to decay into a $W$ boson and $b$ quark with a branching fraction close to one, thus the final states due to $\ttbar$ decay can be classified according to the decay modes of the $W$ bosons.\par
The analysis presented use an integrated luminosity of 1 fb$^{\rm{-1}}$. The $\tau$+lepton analyses were performed on a dataset with an integrated luminosity of 2.2 fb$^{\rm{-1}}$.
\section{DILEPTON FINAL STATES}
In the dilepton final state, the signature of top quark pair decay is two high-$p_T$, opposite charge leptons, missing transverse energy and two or more high-$p_T$ jets~\cite{Arnoud2007b}. The background is due dominantly to $Z/\gamma^*$+jets events, with a small contribution from $WW/WZ/ZZ$ production with additional jets. Both backgrounds are estimated from Monte Carlo simulations. Instrumental backgrounds include fake isolated leptons, either from a jet which is misidentified as an election or from the production of heavy hadrons which decay into leptons, and $\met$ from misidentified jets, calorimeter noise or finite detector resolution.\par
 In the $ee$ channel, we select two electrons with $p_T>15\ \gev$ outside of the $Z$ boson peak, $80\ \gev < M_{ee} < 100\ \gev$. Furthermore, $M_{ee} > 15\ \gev$ and two jets with a transverse momentum of at least 20 GeV are required. The event selection in the dielectron channel is concluded by requiring $\met>40\ \gev$ if $15\ \gev<M_{ee}<80\ \gev$ or $\met>35\ \gev$ if $M_{ee}>100\ \gev$, and sphericity $S>0.15$. Here, $S$ is defined as $S=\frac{3}{2}(\epsilon_1+\epsilon_2)$ where $\epsilon_i$ is the $i^{\rm{th}}$ eigenvalue of the momentum tensor calculated using all electrons and jets. The cross-section in the $ee$ channel was found to be \mbox{$\sigma_{\ppbar\to\ttbar} = 9.6^{+3.2}_{-2.7}\;\rm{(stat)}^{+1.9}_{-1.6}\;\rm{(syst})\pm0.6\;\rm{(lumi)}$}~pb~\cite{Arnoud2007b}.\par
\begin{figure*}[t]
\begin{center}
\subfigure{
\includegraphics[width=0.23\textwidth]{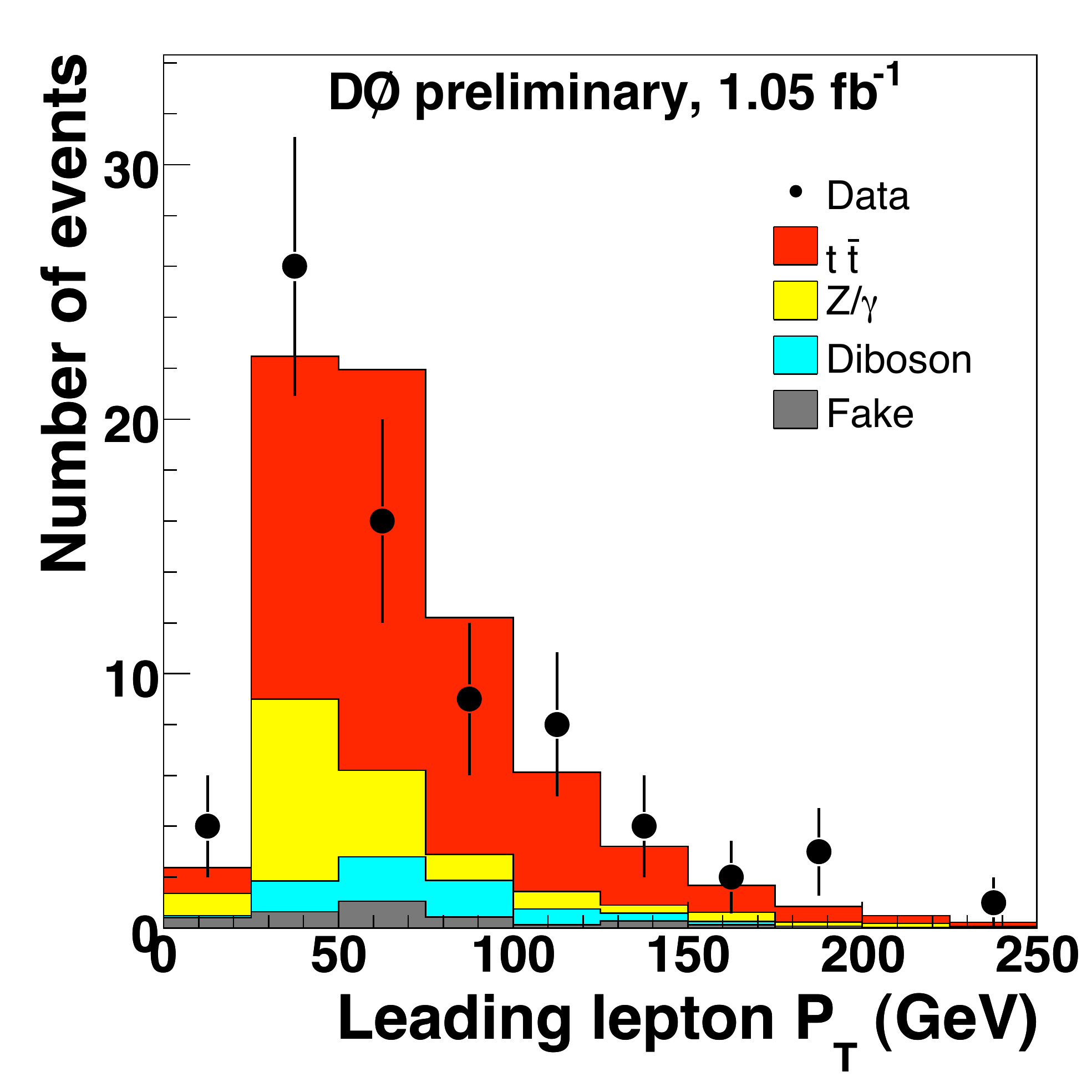}
}
\subfigure{
\includegraphics[width=0.23\textwidth]{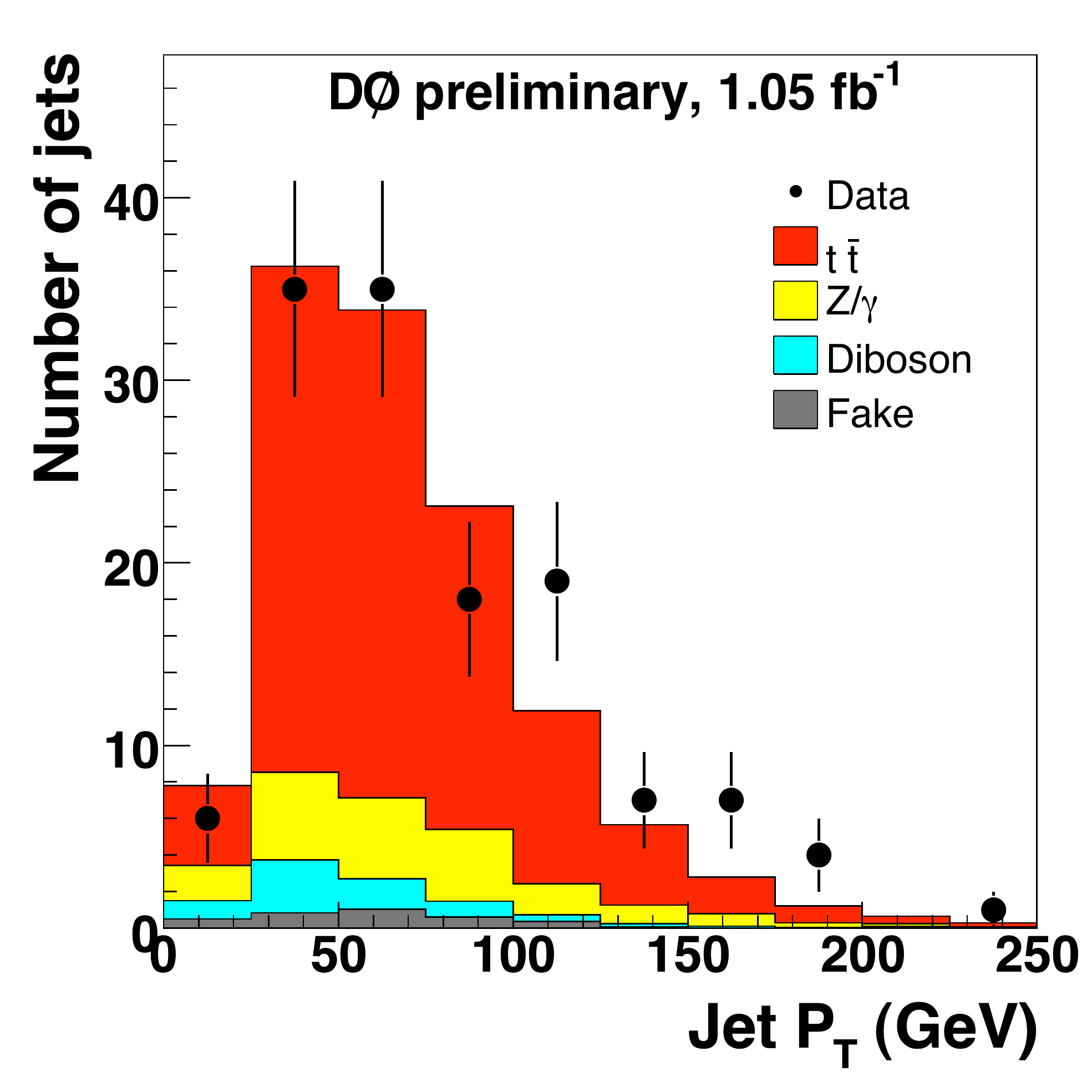}
}
\subfigure{
\includegraphics[width=0.23\textwidth]{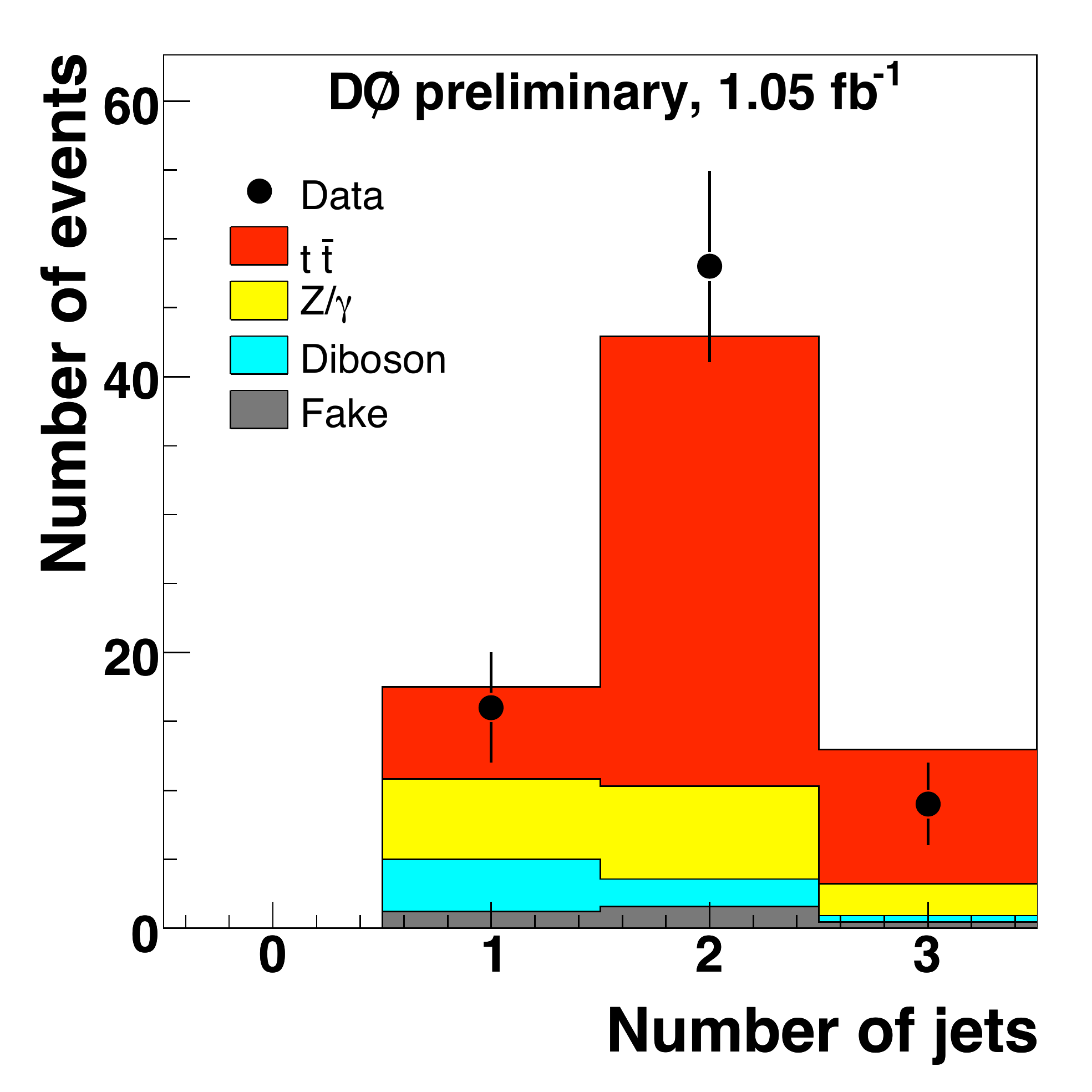}
}
\subfigure{
\includegraphics[width=0.23\textwidth]{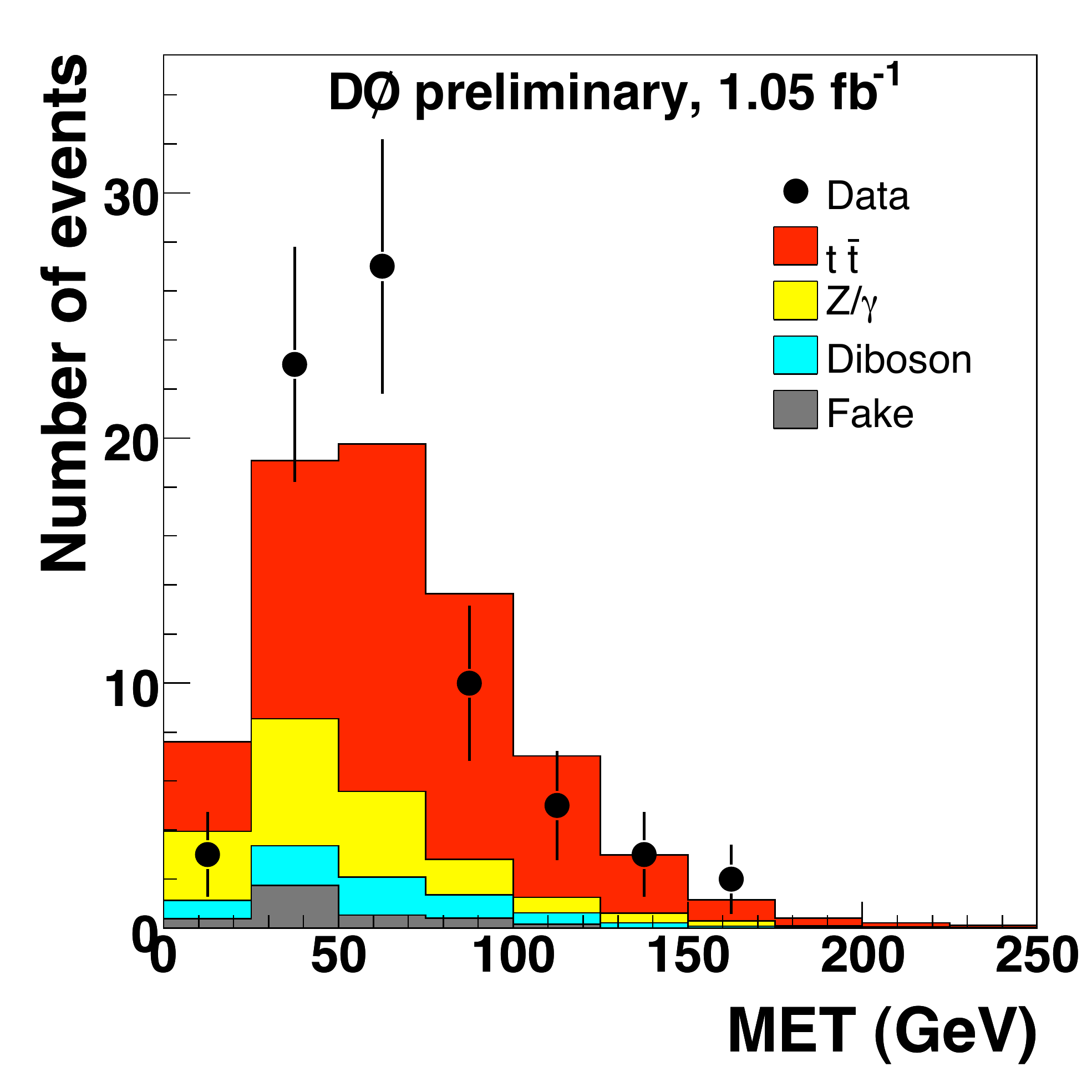}
}
\caption{\label{fig:dilep}Data to Monte Carlo comparison of the leading lepton and jet $p_T$, the jet multiplicity and the $\met$ after the event selection in the dilepton final state.}
\end{center}
\end{figure*}

The event selection in the $e\mu$ final state starts with selecting one electron and one muon with $p_T>15\ \gev$ each. The opposite charge lepton pair with the highest $p_T$ sum is chosen. As in the dielectron channel, we require at least two jets with $p_T>20\ \gev$. Finally, a cut on $H_T$, the scalar sum of the jet transverse momenta of the leading lepton and the two leading jets. We require $H_T>115\ \gev$. In addition to selecting events with two or more jets, we select events with exactly one jet, and require $H_T>105\ \gev$. In Fig. \ref{fig:dilep}, the Monte Carlo prediction of several variables is compared to the observation in data. The cross-section is then measured for both samples, and combined to give \mbox{$\sigma_{\ppbar\to\ttbar} = 6.1^{+1.4}_{-1.2}\;\rm{(stat)}^{+0.8}_{-0.7}\;\rm{(syst})\pm0.4\;\rm{(lumi)}$}~pb~\cite{Arnoud2007b}.\par
In the $\mu\mu$ channel, we select two isolated muons with $p_T>15\ \gev$ and at least two jets with $p_T>20\ \gev$. Furthermore, we require $M_{\mu\mu}>30\ \gev$ and $\met>35\ \gev$. To further reject the $Z$ background, we perform a fit to the invariant dimuon mass and cut on the resulting $\chi^2$. A contour cut in the $\met$ - $\Delta\phi(\met,\ leading\ lepton)$ plane concludes the event selection. In the $\mu\mu$ channel, we measure a cross-section of \mbox{$\sigma_{\ppbar\to\ttbar} = 6.5^{+4.0}_{-3.2}\;\rm{(stat)}^{+1.1}_{-0.9}\;\rm{(syst)}\pm0.4\;\rm{(lumi)}$}~pb~\cite{Arnoud2007b}. The cross-sections measured in the three final states have been combined to give \mbox{$\sigma_{\ppbar\to\ttbar} = 6.8^{+1.2}_{-1.1}\;\rm{(stat)}^{+0.9}_{-0.8}\;\rm{(syst)}\pm0.4\;\rm{(lumi)}$}~pb~\cite{Arnoud2007b}.
\section{LEPTON+JETS FINAL STATES}
For top quark pair decays into lepton+jets, the signature is a high-$p_T$ lepton, large $\met$ and four or more high-$p_T$ jets~\cite{Abazov2008}. The dominating SM background is $W$ production with associated jets, the instrumental background is due to fake isolated leptons in multijet events. To separate signal and background either the kinematic properties of the events are used or b-jet tagging is required.\par
Both analyses start with selecting a muon or electron with $p_T>20\ \gev$ and at least three jets with $p_T>20\ \gev$. For the leading jet, we require $p_T>40\ \gev$, and the $p_T$ sum of the jets has to be greater than 120 GeV if only three jets were found. In the $\mu$+jets channel the missing transverse energy has to be greater than 25 GeV, and the plane angle $\Delta \phi$ between the lepton and the $\met$ has to be greater than $2.1-0.035\met$, while in the $e$+jets channel we require $\met>20\ \gev$ and $\Delta \phi>0.7\pi-0.045\met$.\par
The first analysis employs a topological likelihood to separate between the $\ttbar$ signal and $W$+jets and the multijet background. The topological likelihood is built from up to six kinematic variables that show the best separation between signal and background, and are well modeled by our simulations. The cross-section was measured in the $e$+jets and $\mu$+jets final state separately~\cite{Abazov2008} to give \mbox{$\sigma_{\ppbar\to\ttbar} =  6.3^{+1.0}_{-1.0}\;\rm{(stat)}^{+0.4}_{-0.4}\;\rm{(syst})\pm0.4\;\rm{(lumi)}$}~pb and \mbox{$\sigma_{\ppbar\to\ttbar} = 7.1^{+1.2}_{-1.2}\;\rm{(stat)}^{+0.6}_{-0.5}\;\rm{(syst})\pm0.4\;\rm{(lumi)}$}~pb. Both measurements were combined to yield~\cite{Abazov2008} \mbox{$\sigma_{\ppbar\to\ttbar} = 6.6^{+0.8}_{-0.8}\;\rm{(stat)}^{+0.4}_{-0.4}\;\rm{(syst})\pm0.4\;\rm{(lumi)}$}~pb.\par
\begin{figure*}[t]
\subfigure{
\includegraphics[width=0.24\textwidth]{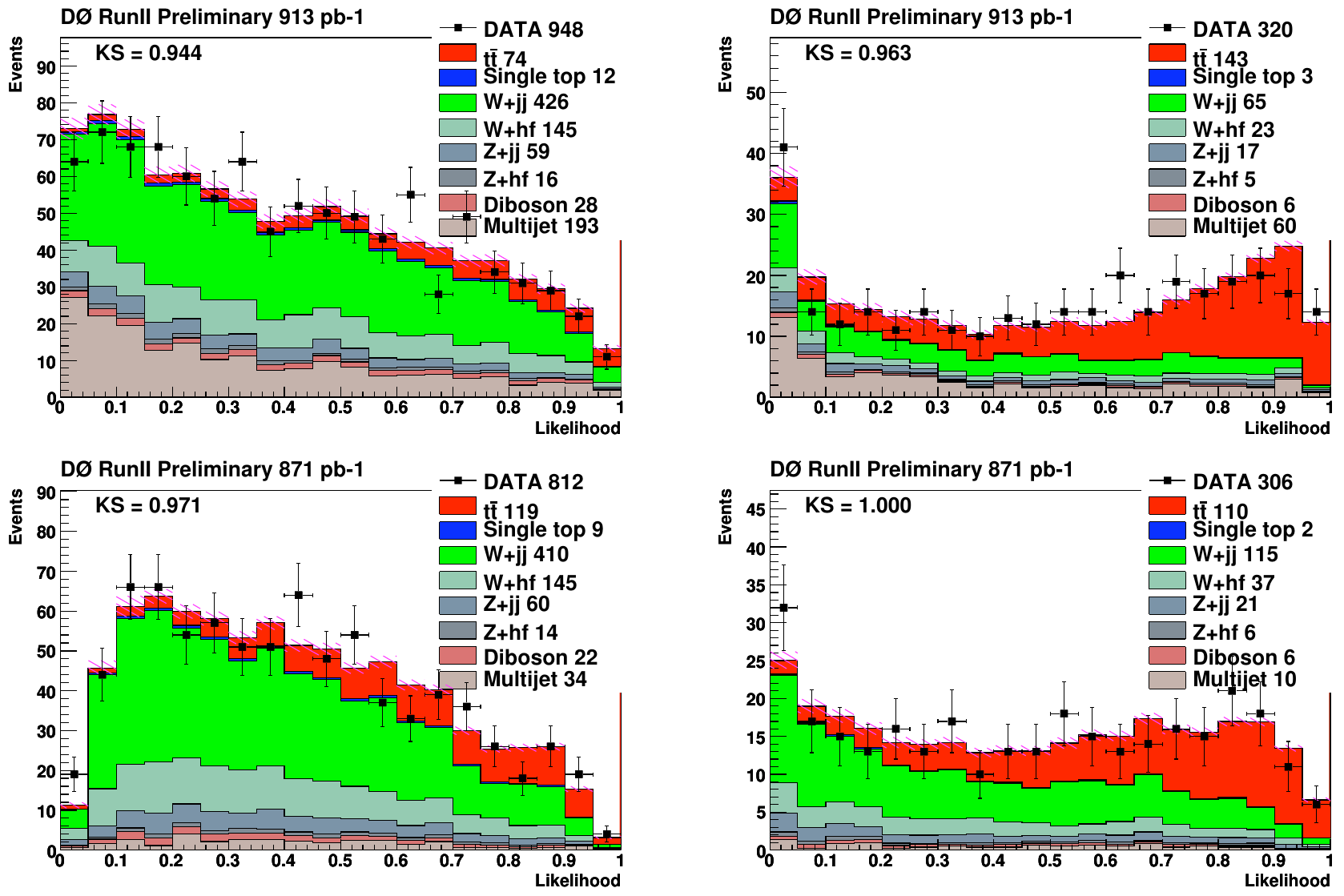}
}
\subfigure{
\includegraphics[width=0.24\textwidth]{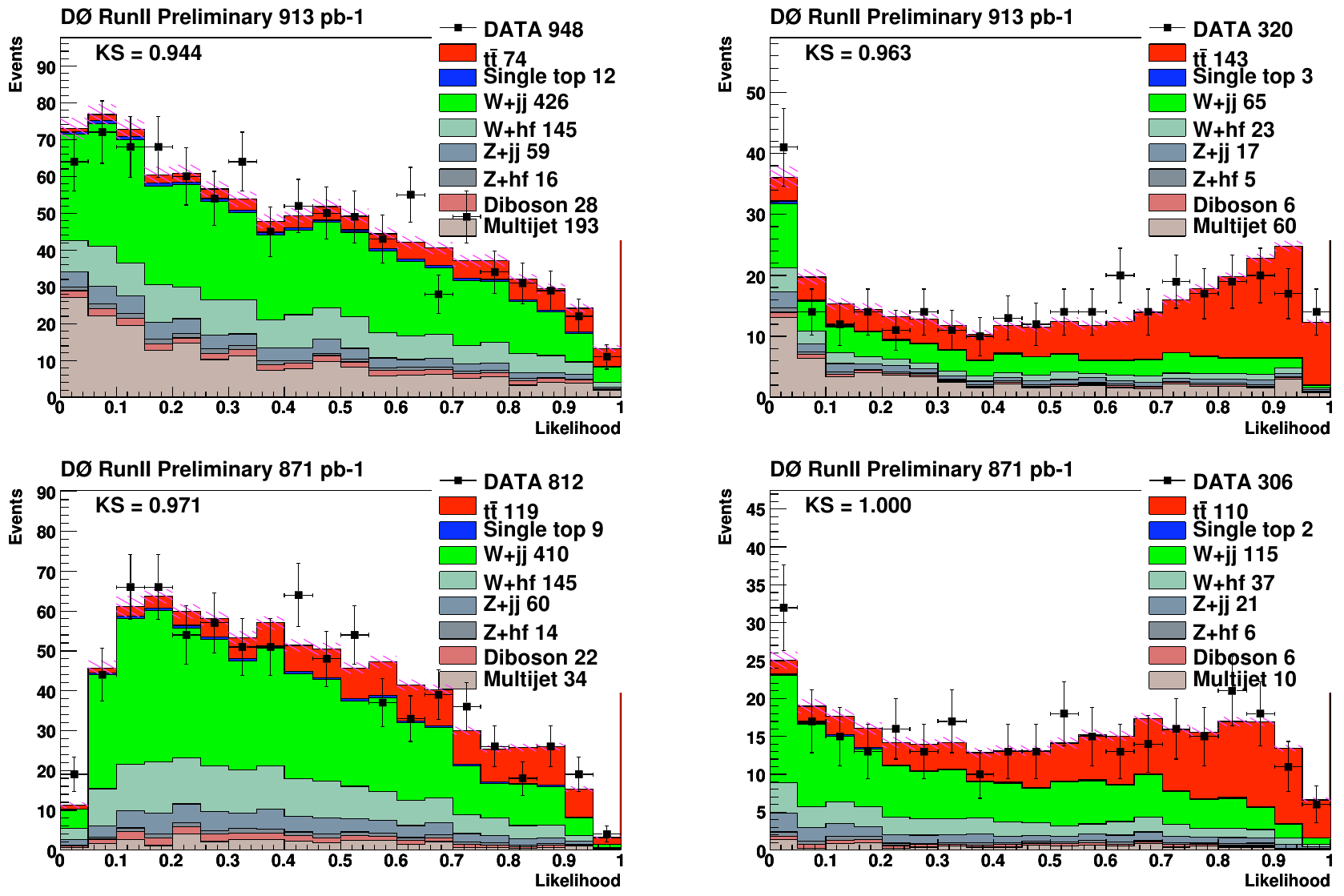}
}
\subfigure{
\includegraphics[width=0.22\textwidth]{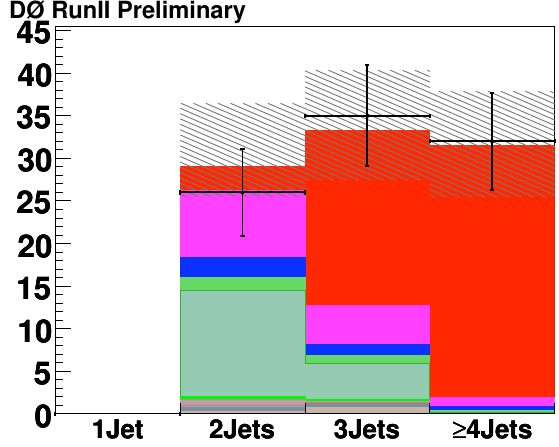}
}
\subfigure{
\includegraphics[width=0.22\textwidth]{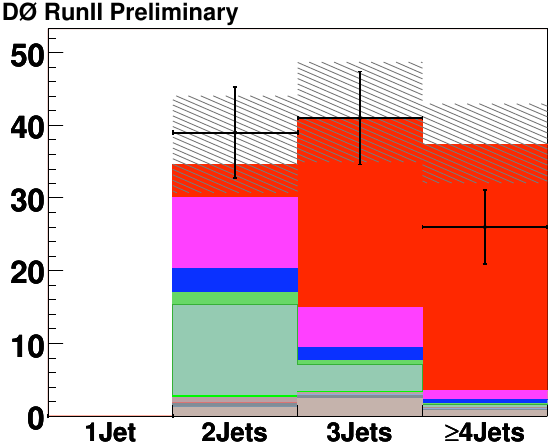}
}
\caption{\label{fig:ljets}From left to right: Kinematic likelihood in the $\mu$+jets and $e$+jets channel and jet multiplicity in the $\mu$+jet and $e$+jet channel after requiring at least two b-tags.}
\end{figure*}

Alternatively, requiring one or more jets to result from the fragmentation of a b-quark reduces the $W+jets$ and multijet background significantly. Thus, we employ a Neural Net (NN) b-tagger to identify b-jets and require either exactly one or two or more b-tags. The cross-section is measured in each b-tag multiplicity bin and combined afterwards. Again, the measurements of the $e$+jets and $\mu$+jets final states, \mbox{$\sigma_{\ppbar\to\ttbar} =  7.3^{+0.7}_{-0.7}\;\rm{(stat)}^{+0.7}_{-0.6}\;\rm{(syst})\pm0.4\;\rm{(lumi pb)}$} and \mbox{$\sigma_{\ppbar\to\ttbar} =  9.0^{+0.9}_{-0.9}\;\rm{(stat)}^{+0.8}_{-0.8}\;\rm{(syst})\pm0.6;\rm{(lumi pb)}$}, have been combined~\cite{Abazov2008} to \mbox{$\sigma_{\ppbar\to\ttbar} = 8.1^{+0.6}_{-0.5}\;\rm{(stat)}^{+0.7}_{-0.7}\;\rm{(syst)}\pm0.5\;\rm{(lumi)}$}~pb. Fig. \ref{fig:ljets} depicts the kinematic likelihood on the left-hand and the jet multiplicity after b-tagging on the right-hand side.
Both the likelihood and b-tagging cross-section measurements have been combined, \mbox{$\sigma_{\ppbar\to\ttbar} = 7.4\pm0.5\;\rm{(stat)}\pm0.5\;\rm{(syst)}\pm0.5\;\rm{(lumi)}$}~pb.
\section{LEPTON+TAU FINAL STATES}
Top quark pairs decaying into a tau lepton and either a muon or an electron have an event signature similar to that of top quark pairs decaying to two light leptons, but one such lepton is now replaced by a hadronically decaying $\tau$. The main SM backgrounds are $W$+jets production and $Z\to\tau\bar{\tau}$, and multijet events with one misidentified light lepton. The analysis is performed on data sets with 1 fb$^{-1}$~\cite{Collaboration2007} and 1.2 fb$^{-1}$~\cite{Collaboration2008} separately with a slightly different selection.\par
\begin{figure*}[t]
\subfigure{
\includegraphics[width=0.23\textwidth]{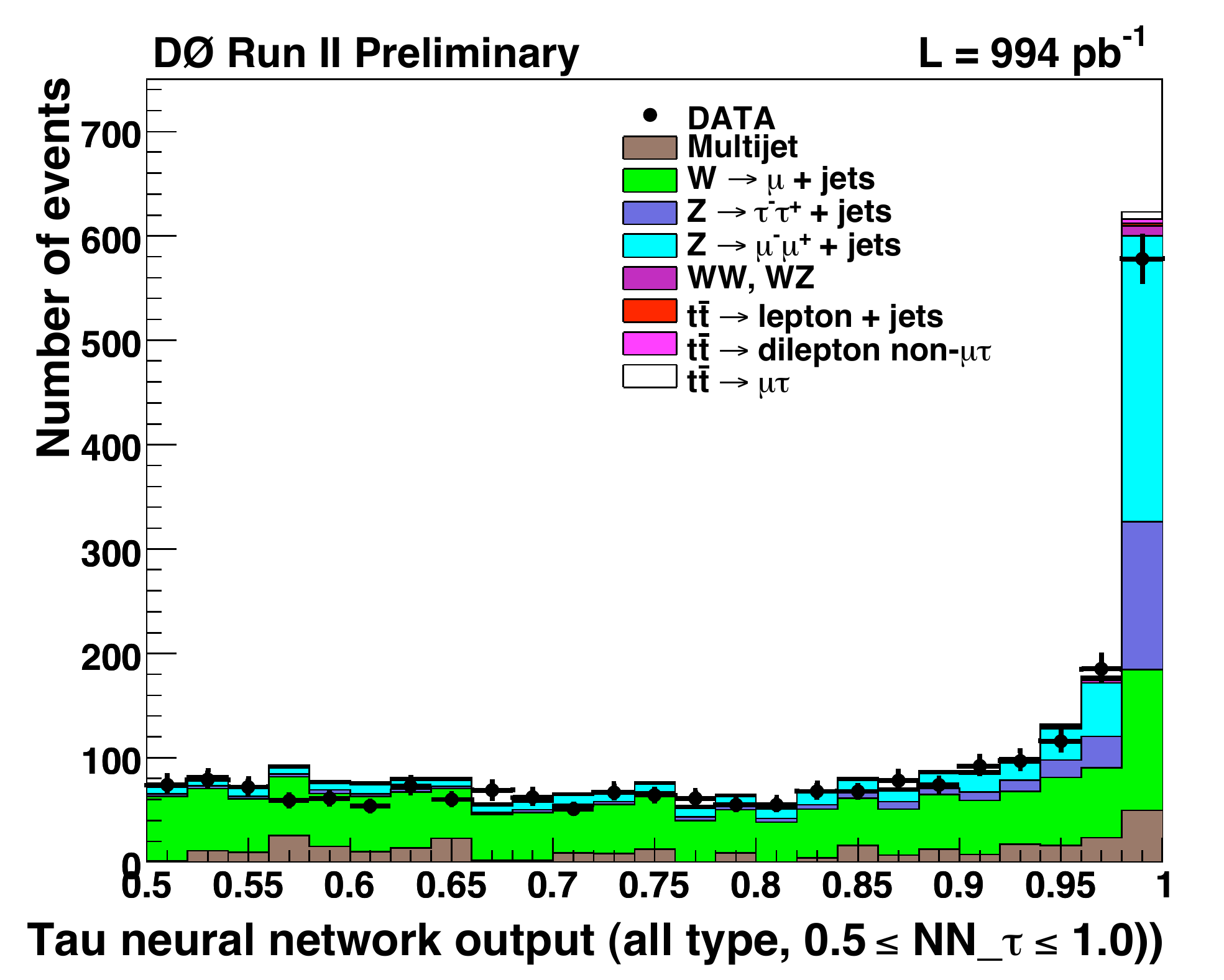}
}
\subfigure{
\includegraphics[width=0.23\textwidth]{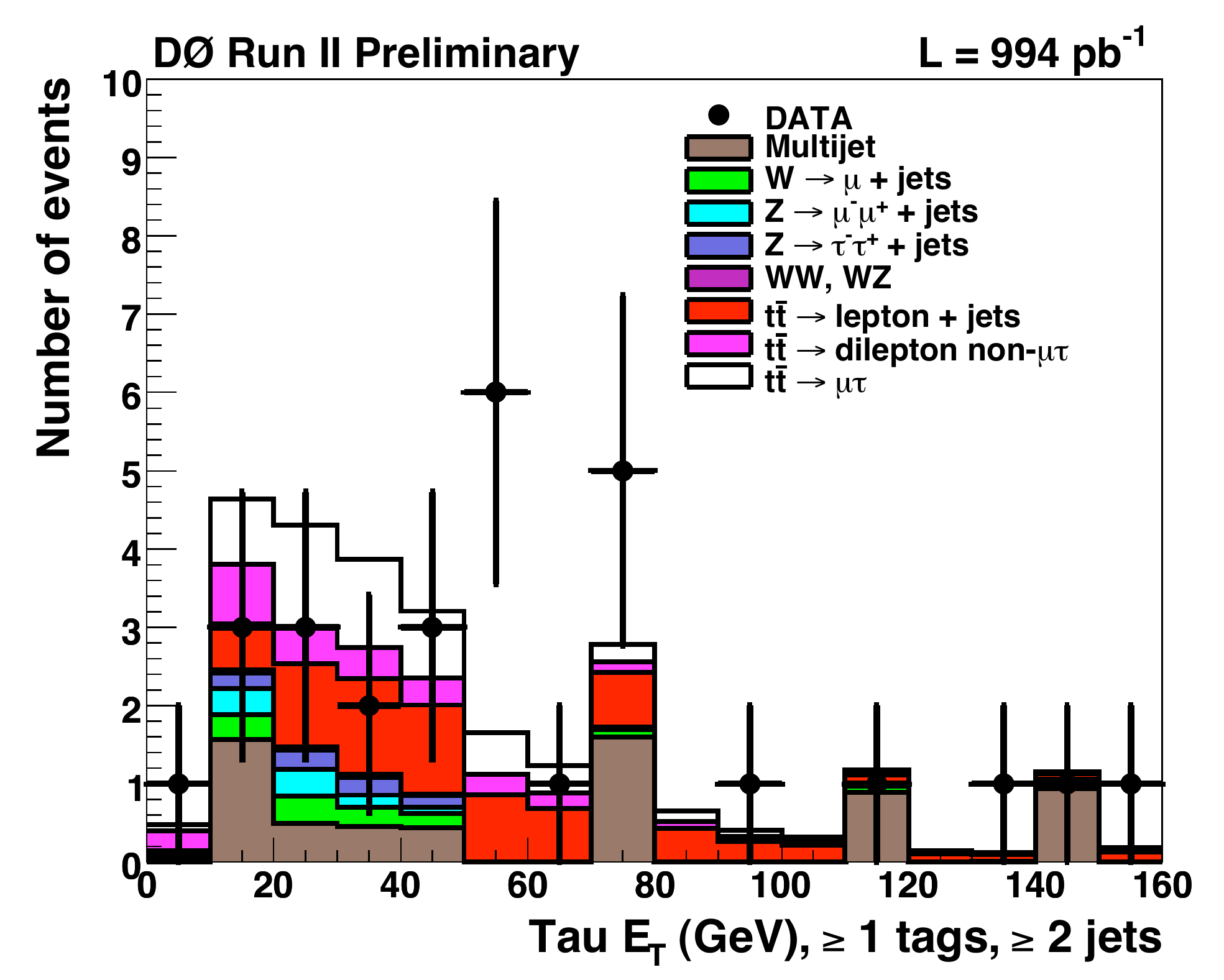}
}
\subfigure{
\includegraphics[width=0.23\textwidth]{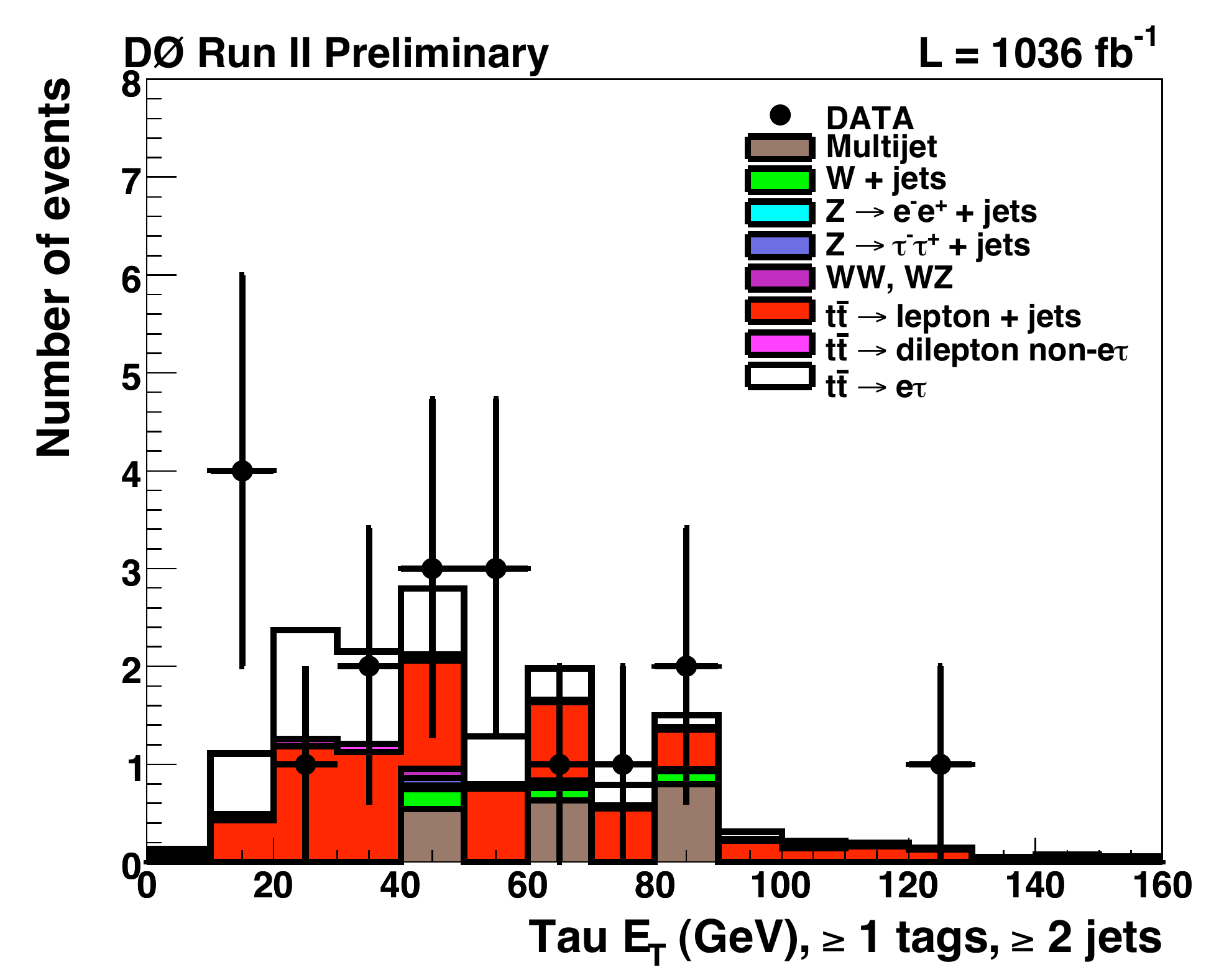}
}
\subfigure{
\includegraphics[width=0.23\textwidth]{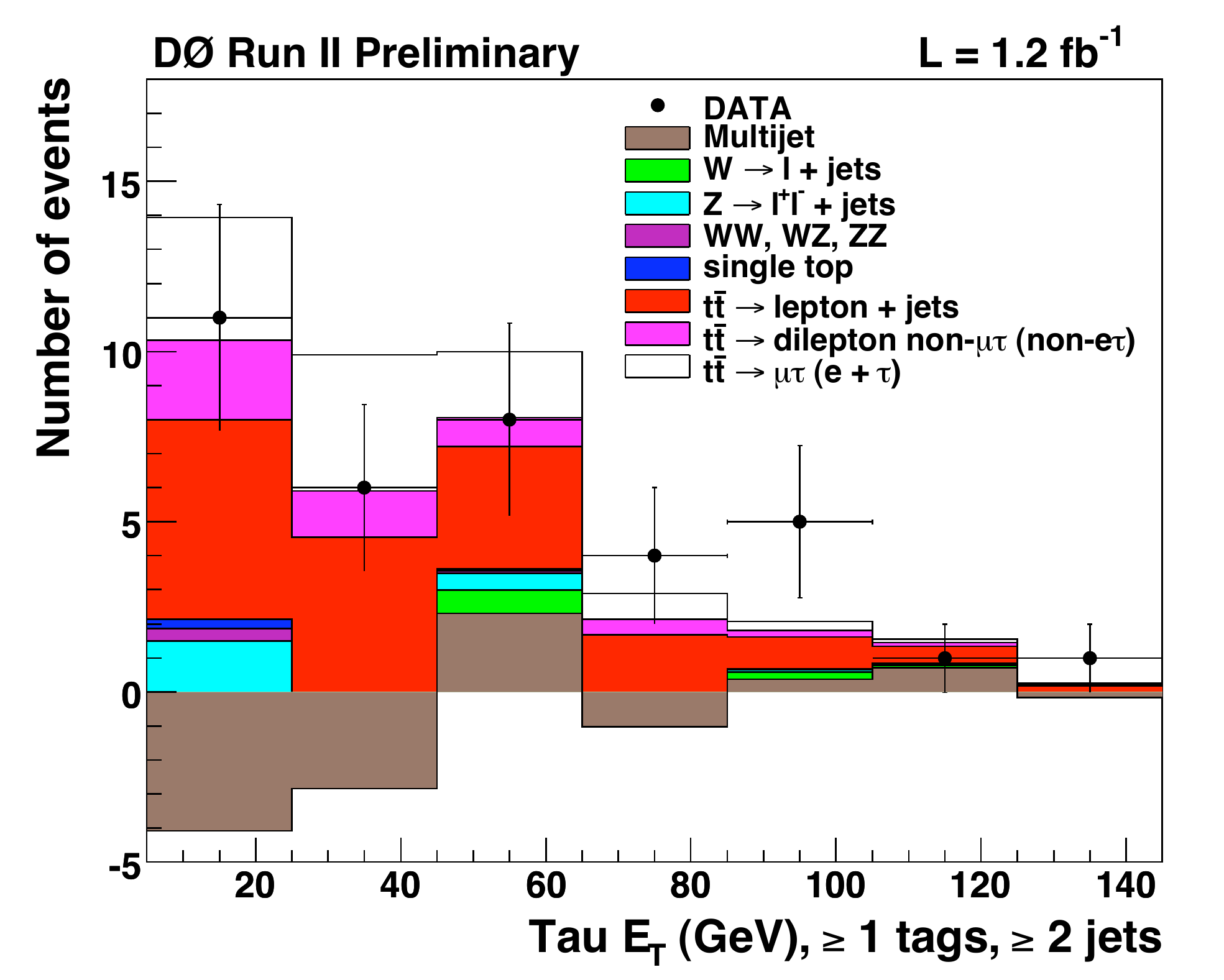}
}
\caption{\label{fig:taujets}From left to right: Output of NN$_{\tau}$, $\tau$ $p_T$ in the $e\tau$ and $\mu\tau$ channels in the 1 fb$^{-1}$ analyses and $\tau$ $p_T$ in the 1.2 fb$^{-1}$ $\ell\tau$ analyses.}
\end{figure*}

Real taus are separated from jets that fake taus by a neural network (NN$_{\tau}$) that uses variables which emphasize the differences between jets and taus such as shower shape and track multiplicity. A second NN is used to separate taus and electrons. The NN$_{\tau}$ output and $\tau$ $p_T$ are shown in Fig. \ref{fig:taujets}.\par
The selection criteria for the lepton+$\tau$ channel consist of either exactly one electron with $p_T>15\ \gev$ or exactly one isolated muon with $p_T>20\ \gev$, at least one jet with $p_T>30\ \gev$ and $15\ \gev < \met < 200\ \gev$. Any additional jet has to have $p_T>20\ \gev$. At least one tau candidate is required. If more than one $\tau$ is found, the one with the highest NN$_\tau$ is used. Jets overlapping with the tau candidate are removed.\par
For the 1 fb$^{-1}$ dataset, a cut on the angle between the electron and $\met$ concludes the selection in the $e\tau$ channel. In the $\mu\tau$ channel, events with a second isolated muon are rejected if the dimuon invariant mass lies in the range $70\ \gev < M_{\mu\mu} < 100\ \gev$. In both final states, at least one b-tag was required to suppress the $W$+jets background. The cross-sections were measured to yield \mbox{$\sigma_{\ppbar\to\ttbar} =  9.6^{+3.2}_{-2.7}\;\rm{(stat)}^{+1.9}_{-1.6}\;\rm{(syst})\pm0.6\;\rm{(lumi)}$}~pb and \mbox{$\sigma_{\ppbar\to\ttbar} =  6.1^{+1.4}_{-1.2}\;\rm{(stat)}^{+0.8}_{-0.7}\;\rm{(syst})\pm0.4\;\rm{(lumi)}$}~pb in the $e\tau$ and $\mu\tau$ final state, respectively. The combination of both final states gives a cross-section of \mbox{$\sigma_{\ppbar\to\ttbar} =  6.8^{+1.2}_{-1.1}\;\rm{(stat)}^{+0.9}_{-0.8}\;\rm{(syst})\pm0.4\;\rm{(lumi)}$}~pb.\par
The selection in the 1.2 fb$^{-1}$ dataset is concluded with a cut on $\met$ and the angle between the lepton and the $\met$ to reject multijet backgrounds, and requiring at least one b-tag. The $\tau$ $p_T$ after b-tagging is depicted in the rightmost plot of Fig \ref{fig:taujets}. The cross-sections in the $e\tau$ and $\mu\tau$ final state were determined to be \mbox{$\sigma_{\ppbar\to\ttbar} = 9.6^{+3.2}_{-2.7}\;\rm{(stat)}^{+1.9}_{-1.6}\;\rm{(syst})\pm0.6\;\rm{(lumi)}$}~pb and \mbox{$\sigma_{\ppbar\to\ttbar} =  6.1^{+1.4}_{-1.2}\;\rm{(stat)}^{+0.8}_{-0.7}\;\rm{(syst})\pm0.4\;\rm{(lumi)}$}~pb. Their combination yields \mbox{$\sigma_{\ppbar\to\ttbar} =  6.8^{+1.2}_{-1.1}\;\rm{(stat)}^{+0.9}_{-0.8}\;\rm{(syst})\pm0.4\;\rm{(lumi)}$}~pb.\par
The combination of the cross-sections measured in both datasets is $\sigma_{\ppbar\to\ttbar} =  8.1^{+0.6}_{-0.5}\;\rm{(stat)}^{+0.7}_{-0.7}\;\rm{(syst})\pm0.5\;\rm{(lumi)}$~pb.
\section{SUMMARY}
\begin{figure*}[t]
\includegraphics[width=0.5\textwidth]{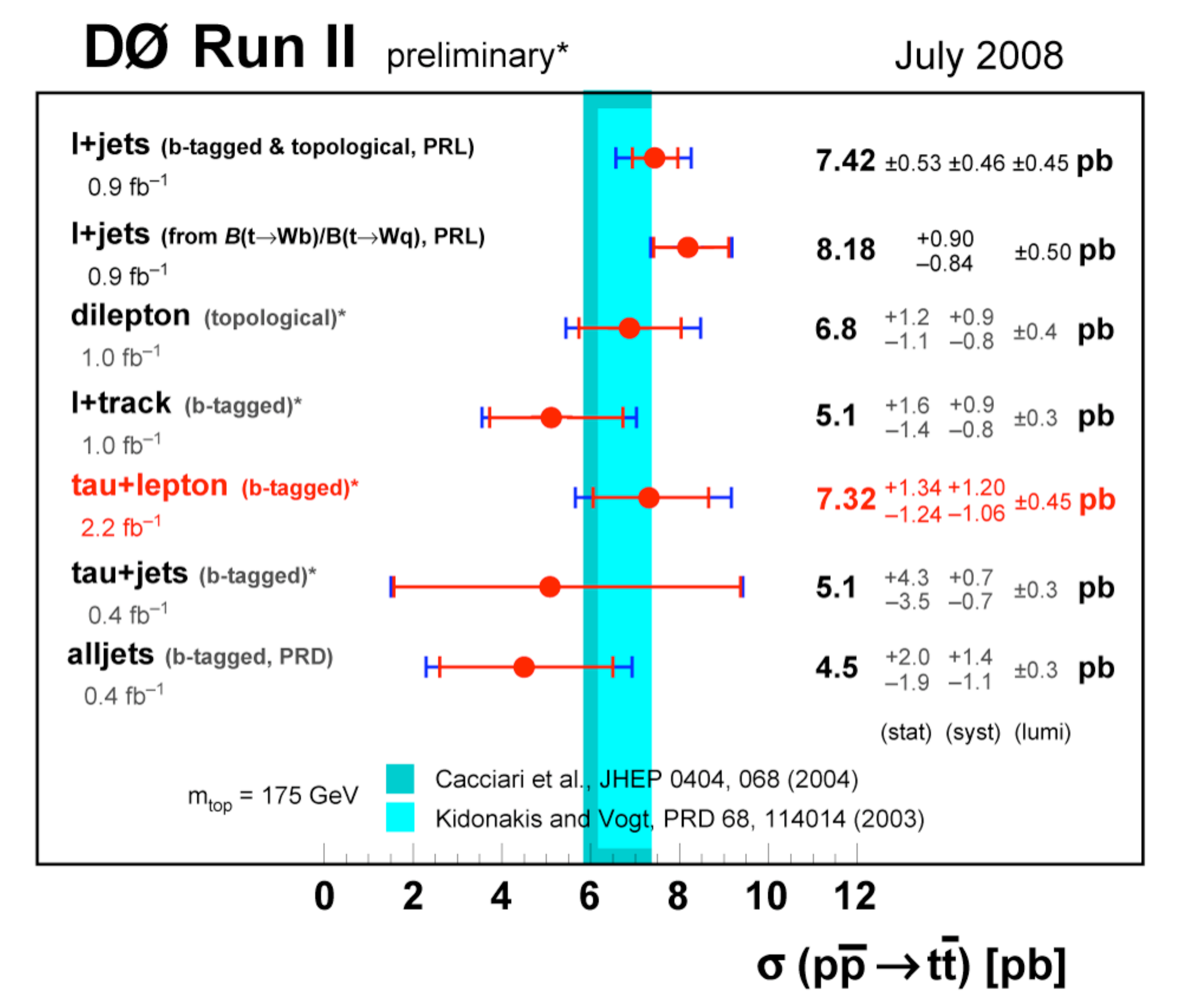}
\caption{\label{fig:xssummary}Graphical summary of top quark cross-section measurements carried out at D\O .}
\end{figure*}

The top quark pair production cross-section has been measured by the D\O\ collaboration in $\ppbar$ collisions with $\sqrt{s}=$1.96 TeV in different final states. The cross-section in the $\ell\tau$ final state was measured for the first time. All cross-sections are compatible within their errors and with the theoretical predictions. Fig. \ref{fig:xssummary} summarizes the cross-section measurements and compared them to two theoretical predictions.
\section{ACKNOWLEDGMENTS}
We thank the staffs at Fermilab and collaborating institutions, and acknowledge support from the DOE and NSF (USA); CEA and CNRS/IN2P3 (France); FASI, Rosatom and RFBR (Russia); CNPq, FAPERJ, FAPESP and FUNDUNESP (Brazil); DAE and DST (India); Colciencias (Colombia); CONACyT (Mexico); KRF and KOSEF (Korea); CONICET and UBACyT (Argentina); FOM (The Netherlands); STFC (United Kingdom); MSMT and GACR (Czech Republic); CRC Program, CFI, NSERC and WestGrid Project (Canada); BMBF and DFG (Germany); SFI (Ireland); The Swedish Research Council (Sweden); CAS and CNSF (China); and the Alexander von Humboldt Foundation (Germany).
\bibliography{literatur}\bibliographystyle{thesis}
\end{document}